\documentclass{article}

\usepackage{arxiv}
\usepackage[utf8]{inputenc} 
\usepackage[T1]{fontenc}    
\usepackage{hyperref}       
\usepackage{url}            
\usepackage{booktabs}       
\usepackage{amsfonts}       
\usepackage{nicefrac}       
\usepackage{microtype}      
\usepackage{lipsum}		    
\usepackage{graphicx}
\usepackage{natbib}
\usepackage{doi}
\usepackage{siunitx}
\usepackage{amsmath,amssymb,amsfonts}
\usepackage{mathtools} 
\usepackage{pifont}
\usepackage{float}
\usepackage{caption}
\usepackage{subcaption}
\usepackage{comment}

\usepackage{multirow}
\usepackage{color}

\title{Toward High Performance, Programmable Extreme-Edge Intelligence for Neuromorphic Vision Sensors utilizing Magnetic Domain Wall Motion-based MTJ}

\author{Md Abdullah-Al Kaiser\thanks{These authors contributed equally to this work.} \\
	University of Southern California\\
	Los Angeles, CA \\
	\texttt{mdabdull@usc.edu} \\
	\And
	Gourav Datta$^*$ \\
	University of Southern California\\
	Los Angeles, CA \\
	\texttt{gdatta@usc.edu} \\
        \And
	Peter A. Beerel \\
	University of Southern California\\
	Los Angeles, CA  \\
	\texttt{pabeerel@usc.edu} \\
 	\And
	Akhilesh R. Jaiswal \\
	University of Wisconsin-Madison\\
	Madison, WI  \\
	\texttt{akhilesh.jaiswal@wisc.edu} \\
}
\renewcommand{\shorttitle}{Neuromorphic CMOS+X P\si{^2}M}

\begin{document}
\maketitle

\begin{abstract}
The desire to empower resource-limited edge devices with computer vision (CV) must overcome the high energy consumption of collecting and processing vast sensory data. To address the challenge, this work proposes an energy-efficient non-von-Neumann in-pixel processing solution for neuromorphic vision sensors employing emerging (X) magnetic domain wall magnetic tunnel junction (MDWMTJ) for the first time, in conjunction with CMOS-based neuromorphic pixels. Our hybrid CMOS+X approach performs in-situ massively parallel asynchronous analog convolution, exhibiting low power consumption and high accuracy across various CV applications by leveraging the non-volatility and programmability of the MDWMTJ. Moreover, our developed device-circuit-algorithm co-design framework captures device constraints (low tunnel-magnetoresistance, low dynamic range) and circuit constraints (non-linearity, process variation, area consideration) based on monte-carlo simulations and device parameters utilizing GF22nm FD-SOI technology. Our experimental results suggest we can achieve an average of 45.3\% reduction in backend-processor energy, maintaining similar front-end energy compared to the state-of-the-art and high accuracy of 79.17\% and 95.99\% on the DVS-CIFAR10 and IBM DVS128-Gesture datasets, respectively. 
\end{abstract}


\maketitle

\section{Introduction}

Edge devices with computer vision (CV) systems face energy inefficiency and throughput bottlenecks due to physically segregated sensor hardware and processing platforms \cite{system_bottleneck}. To address this, researchers have explored near-sensor, in-sensor, and in-pixel processing methods \cite{near_sensor_3D_sony, sleepspotter, inpix_conv, aps_p2m, mixed_mode_ivs}. While near-sensor and in-sensor approaches still encounter bandwidth challenges, in-pixel processing allows simultaneous sensing and computing within the pixel array, providing energy-efficient processing by transmitting feature outputs instead of raw sensory data. In the realm of CMOS Image Sensors (CIS), bio-inspired event-based neuromorphic vision sensors (NVS) \cite{DVS_ref1, DVS_ref2} have gained popularity over traditional CIS for various neural network (NN) applications \cite{DVS_auto_driving, DVS_pose, DVS_steering} due to their lower latency, energy efficiency, high temporal precision, and dynamic range. In summary, integrating neuromorphic vision sensors with in-pixel processing presents a comprehensive solution to energy inefficiency and throughput bottlenecks in resource-constrained edge devices with CV systems.

Neuromorphic vision sensors often utilize spiking convolutional neural networks (CNNs) for processing asynchronous input events. In the traditional approach, the duration of the neuromorphic datasets is segmented into predetermined integration intervals, accumulating input spikes within each period to generate multi-bit inputs for the initial layer of the spiking CNN \cite{spiking_cnn}. Unlike subsequent layers that consist of energy-efficient accumulators, the first layer involves digital multi-bit Multiply-Accumulate (MAC) operations. To enhance energy efficiency in the first layer, analog MAC units, employing continuous variables like current, resistance, or pulse width as weights and capacitors as accumulators (representing the membrane potential), can be utilized in CNN hardware implementations \cite{sleepspotter, mixed_mode_ivs, reconfig_inpix, inmemory_nvs, current_cap_integ, dvsp2m}. Substantial energy consumption may result from the active amplifier-based capacitor; in contrast, the passive capacitor-based accumulator yields low-energy consumption. However, the passive capacitor cannot retain the charge (membrane potential) for a long duration due to the leakage of the CMOS circuits, resulting in lower overall classification accuracy. 

Due to the limitations of the CMOS circuits including the inherent leakage, research is shifting towards NN architectures utilizing emerging (X) post-CMOS technologies such as Phase Change Memories (PCM), Resistive Random Access Memory (RRAM), Magnetic Tunnel Junction (MTJ), Magnetic Domain Wall (MDW), etc. \cite{inmem_emerging, inmem_rram, mdw_ann, mdw_exp1}. The benefits of non-volatility, reduced power consumption, higher density, speed, and CMOS compatibility drive this shift. Spintronics devices offer lower latency, reduced energy dissipation, and unlimited endurance compared to other emerging technologies; however, they suffer from a low on-off ratio due to low TMR (between 200\% and 600\% \cite{mdw_exp2, tmr_600p}). 
Spintronic device magnetic domain wall magnetic tunnel junction (MDWMTJ) demonstrates a continuous resistance state based on domain wall position and has experimentally demonstrated good accuracy in neuromorphic applications \cite{mdw_exp1, mdw_exp2}. In addition, hybrid CMOS and MDWMTJ structures are reported for logic, in-memory, and neuromorphic applications \cite{mdw_ann, mdw_logic1, mdwmtj_ref1}. Considering these advantages, our proposed processing-in-pixel-in-memory (P\si{^2}M) hardware for neuromorphic vision sensors utilizes MDWMTJ as the core component. Note, we choose spin-orbit-torque (SOT)-based MDWMTJ instead of spin-transfer-torque (STT)-based MDWMTJ due to their lower write current requirements and decoupled read and write path, resulting in low power dissipation and constant resistance along the write path that makes the associated CMOS circuit design easier. 

This work presents an energy-efficient in-pixel processing hardware for spiking CNN focusing on neuromorphic vision applications. We utilize hybrid CMOS and MDWMTJ approaches to achieve high dynamic range, low energy consumption, and programmability for our neuromorphic CMOS+X P\si{^2}M hardware. The key contributions of our work include the following:

\begin{enumerate}
    \item  We propose a novel hybrid CMOS+X approach of processing-in-pixel-in-memory (P\si{^2}M) for neuromorphic vision sensors, incorporating the emerging (X) magnetic domain wall magnetic tunnel junction (MDWMTJ) device as the core compute element for our developed spiking CNN framework. The non-volatility and programmability of the MDWMTJ enable the membrane potential to be retained for a longer integration time and tunable weight and neuron threshold, which are important for ensuring high accuracy. 
    \item In addition, we design three current-based analog weight configurations: CMOS-based, MDWMTJ-based, and hybrid CMOS+X. Our hybrid symbiotic CMOS+X approach combines the unique benefits of both technologies, exhibiting close to state-of-the-art (SOTA) accuracy (thanks to the high dynamic range supported by the CMOS weights) while allowing mapping of multiple CV applications onto the same hardware (owing to the programmability of MDWMTJ).
    \item Finally, we develop a device-circuit-algorithm co-design solution incorporating the device constraints (low TMR, low dynamic range) and circuit constraints (non-linearity, process variations, area limitations based on the extensive monte-carlo simulations and parameters utilizing GF22nm FD-SOI technology node) into our algorithmic framework, resulting in an average of 45.3\% reduction in backend-processor energy consumption on the NMNIST, CIFAR10-DVS, IBM DVS128-Gesture dataset, albeit with a 0.28\%, 1.55\%, and 1.23\% in test accuracy drop from the baseline, respectively.
\end{enumerate}

To the best of our knowledge, this is the first work to present a pathway towards attaining \textit{high-accuracy and programmability on complex neuromorphic datasets} using in-pixel processing for NVS cameras leveraging a CMOS+X solution. 

\section{Device Preliminaries and Modeling}

\begin{figure}[!t]
\centering
\subfloat[]{\includegraphics[width=0.70\linewidth]{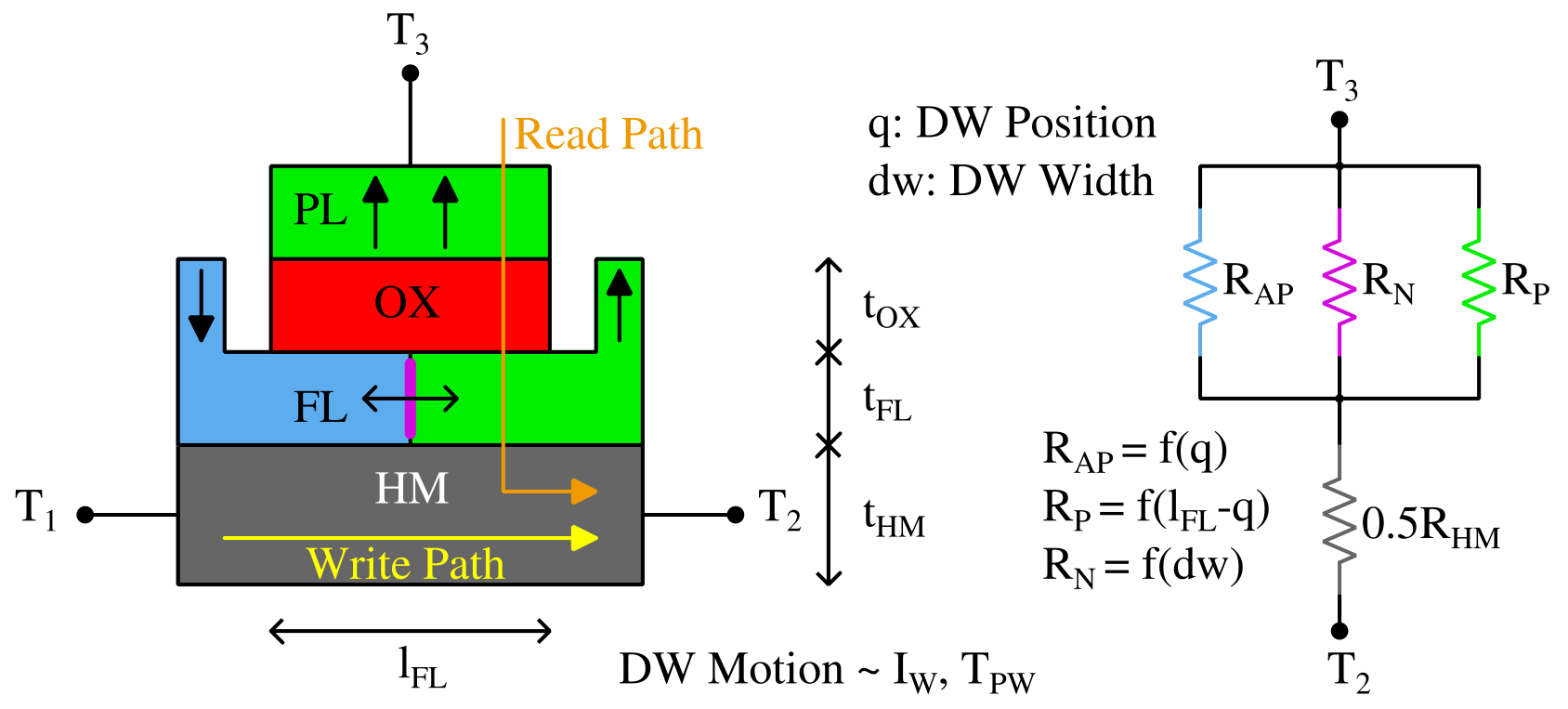}} \\
\subfloat[]{\includegraphics[width=0.33\linewidth]{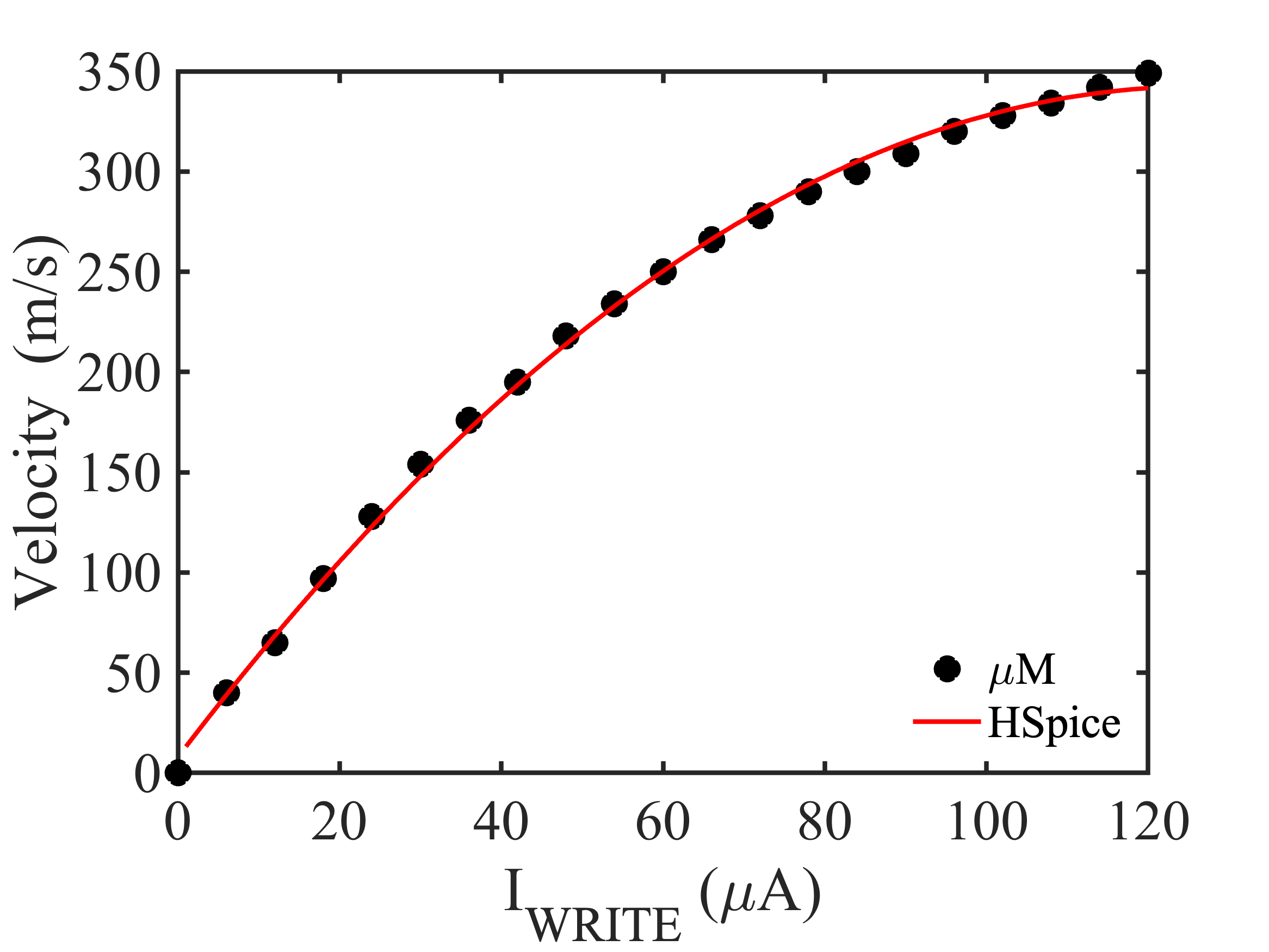}}
\subfloat[]{\includegraphics[width=0.33\linewidth]{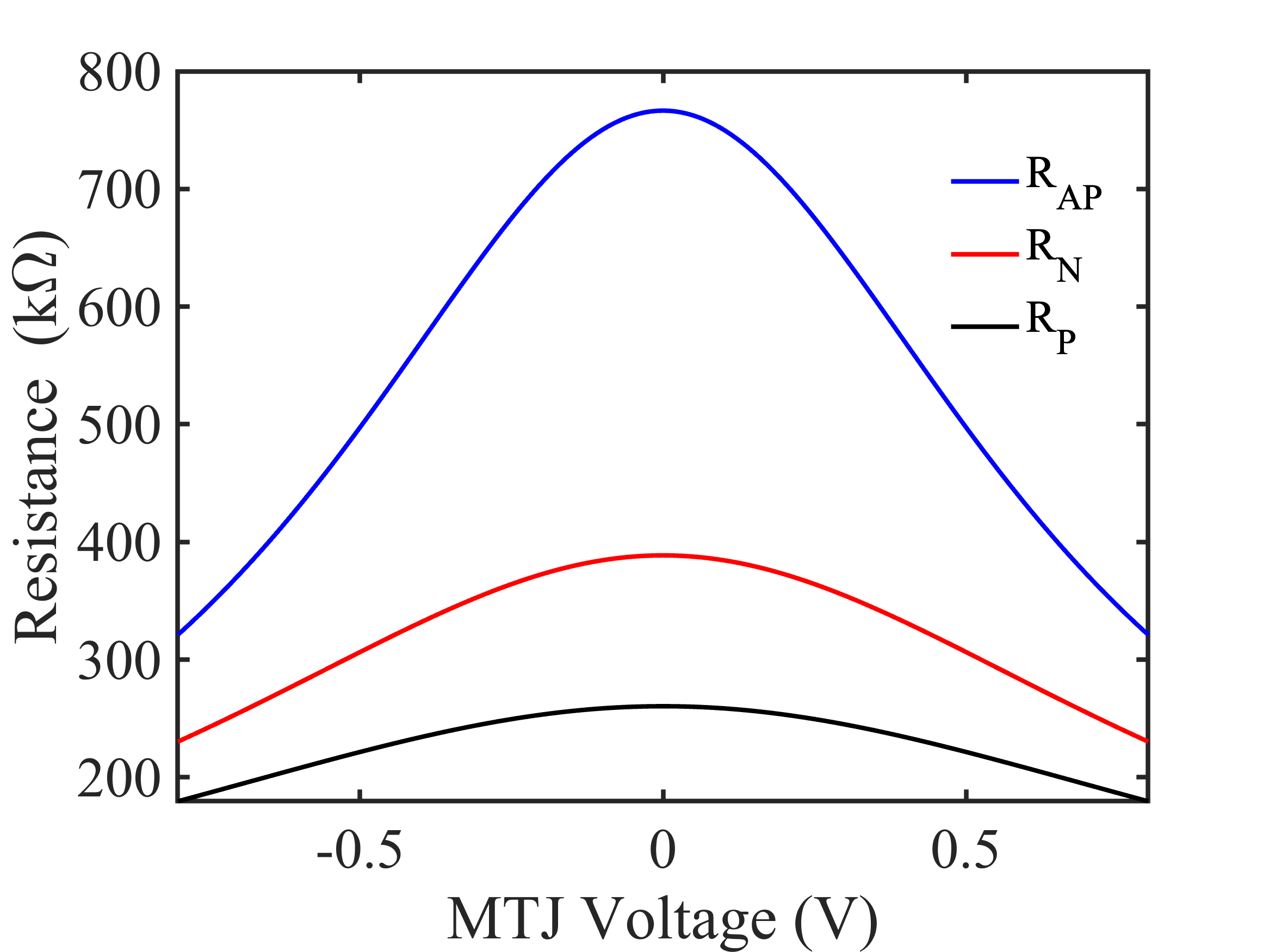}}
\subfloat[]{\includegraphics[width=0.33\linewidth]{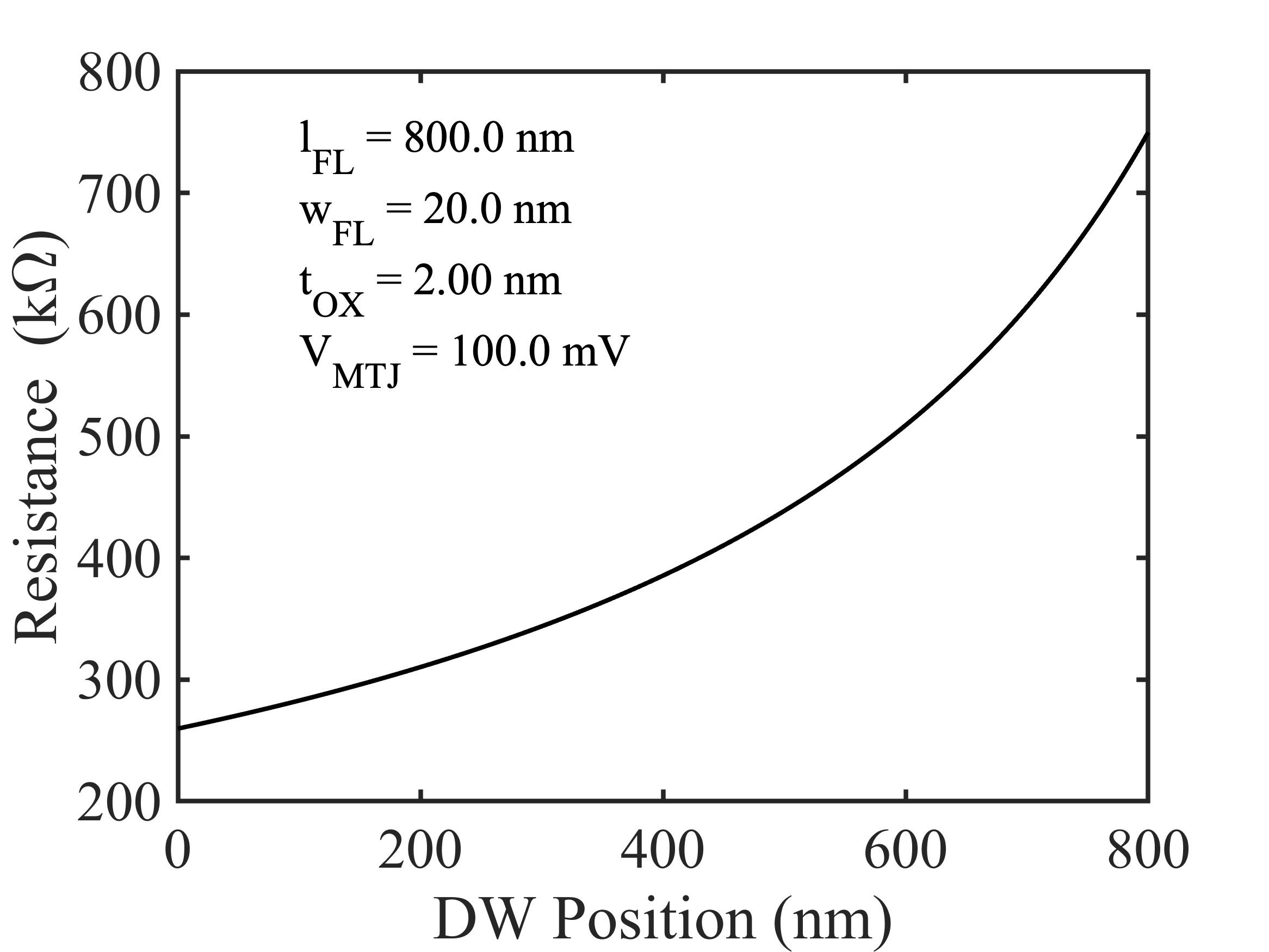}}
\caption{(a) Device structures for SOT-based MDWMTJ. \si{l_{FL}} \si{t_{FL}} denote the length and thickness of the FL, \si{t_{OX}} and \si{t_{HM}} represent the thickness of the oxide and HM layer, respectively. \si{R_{P}}, \si{R_{AP}}, and \si{R_{N}} denote the MTJ resistance of the parallel, and-parallel and perpendicular state. (b) DW velocity as a function of the write current. (c) MTJ resistance as a function of applied voltage. (d) Effective MDWMTJ resistance as a function of the DW position.} 
\label{device_modeling}
\end{figure}

Figure \ref{device_modeling}(a) depicts the structure of a spin-orbit-torque (SOT)-based MDWMTJ. It consists of a thin insulating oxide layer between two ferromagnetic (FM) layers: a pinned layer (PL) with fixed magnetization and a free layer (FL) with parallel (P) and antiparallel (AP) magnetic domains. The AP domain has higher resistance due to the tunnel magnetoresistance (TMR) effect. A heavy metal (HM) layer below the FL influences the domain wall (DW) position through current-induced SOT and Dzyaloshinskii-Moriya interaction (DMI) \cite{mdwmtj_ref1, mdwmtj_ref2, dmi_ref1}. Unlike conventional MTJs, the MDW-MTJ exhibits continuous resistance states based on the DW position (q), allowing for multiple resistance levels, as experimentally demonstrated \cite{mdw_exp1}.

In the SOT-based MDWMTJ, the read and write paths are decoupled, allowing independent optimization. The DW position is programmed by applying a current pulse through the HM between the terminal T1 and T2 (Figure \ref{device_modeling}(a)), where the DW motion is proportional to pulse amplitude and duration. Our approach utilizes the T1 to T2 directed (positive direction) current flow for weighted accumulation, a configuration experimentally demonstrated in \cite{mdw_integrator}. A long-duration current from T2 to T1 (negative direction) is applied to reset the DW position (q), which shifts the DW to the leftmost position (q = 0). The MDWMTJ resistance (between T3 and T2) is non-linearly dependent on DW position and applied voltage. At q = 0 and q = \si{l_{FL}}, MDWMTJ exhibits the lowest (\si{R_P}) and highest (\si{R_{AP}}) resistance, respectively. For other DW positions (q), the resistance non-linearly varies between \si{R_P} and \si{R_{AP}}.

We developed a compact Verilog-A model of the MDWMTJ for HSpice simulations, benchmarking it with MuMax3 \si{\mu}M results utilizing experimental device parameters from \cite{mdwmtj_ref1}. The Verilog-A model captures DW velocity as a function of the write current utilizing a 2nd-order polynomial fitting function (Figure \ref{device_modeling}(b)), demonstrating 1.06\% of RMSE error. Using a resistance model that has been benchmarked to a non-equilibrium Green’s function (NEGF)-based framework from \cite{mdw_negf}, we simulated MTJ resistance across different applied voltages (Figure \ref{device_modeling}(c)). To determine the resistance of an MTJ in an FM with a domain wall between two oppositely polarized domains, the NEGF-based simulator is adapted to account for a parallel connection of three separate MTJs (parallel, anti-parallel, and perpendicular) \cite{mdw_ann}. The parallel and anti-parallel resistance varies as a function of the DW position, while the perpendicular resistance depends on the DW width. The effective MDWMTJ resistance, including the HM resistance as a function of the DW position, is shown in Figure \ref{device_modeling}(d). Our analysis utilizes the compact 3-terminal MDWMTJ Verilog-A model, capturing the DW dynamics and non-linear voltage-dependent resistance.

\section{Proposed neuromorphic CMOS+X P\si{^2}M Hardware Architecture}
  
\begin{figure}[!t]
\centering
\includegraphics[width=1\linewidth]{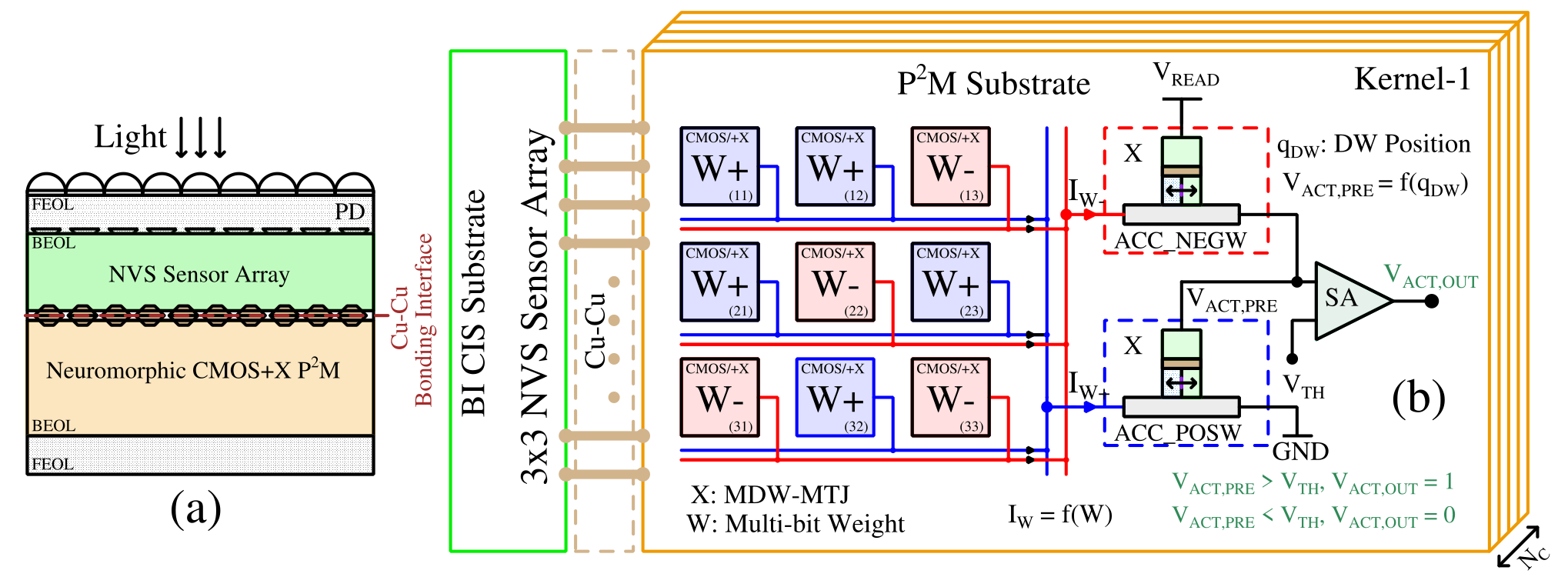}
\caption{(a) The representative 3D heterogeneously integrated CMOS+X P\si{^2}M architecture utilizing Cu-Cu hybrid bonding, where the top die is backside illuminated CIS substrate, and the bottom die consists of P\si{^2}M compute elements. (b) A computing architecture of the MAC and thresholding operation considering a kernel size of 3\si{\times}3.}
\label{dvs_p2m_arch}
\vspace{-0.25in}
\end{figure}

This section introduces the hardware innovations and implementation of our proposed CMOS+X spiking CNN architecture. The in-situ P\si{^2}M hardware can be heterogeneously 3D integrated \cite{3D_ref1, 3D_ref2} where the top die accommodates the NVS array, and the bottom die contains vertically aligned MAC units per the spiking CNN filter inputs  (Figure \ref{dvs_p2m_arch}). The sensor array generates an ON (OFF) event spike when the contrast detected by the neuromorphic pixel increases (decreases) by a certain threshold. The neuromorphic CMOS+X P\si{^2}M core on the bottom die receives event spikes as inputs from the top die through hybrid Cu-Cu bonding. The P\si{^2}M core consists of analog weights and accumulators to perform the MAC computations. The current represents the weight values encoded as device parameters, such as the width of CMOS transistors, or the resistance of the MDWMTJ, or both. The input-current-modulated DW motion of the MDWMTJ has been utilized as the accumulator.

Spiking CNN requires asynchronous MAC computation and synchronous thresholding after a fixed integration time. Input event spikes trigger asynchronous write currents through the accumulator MDWMTJ, dependent on weight values. For large (small) weight values, large (small) write current (\si{I_W}) flows through the HM, resulting in large (small) DW motion from its previous state (position). As the input spikes are binary,  DW motion represents the accumulated MAC output, with separate MDWMTJs for positive (\si{ACC\_POSW}) and negative (\si{ACC\_NEGW}) weights. After the integration time, the pre-activation voltage (\si{V_{ACT,PRE}}), non-linearly proportional to DW position differences of the positive and negative accumulators, is generated. Finally, a thresholding circuit compares the pre-activation voltage with the threshold, producing the output activation spike for the next layer if it exceeds the threshold. A tunable threshold per neuron is achieved using a series divider of two programmable MDWMTJs, which will be utilized in the algorithmic optimization, along with the programmable weights for catering to various applications. 

The spiking CNN requires multiple channels in the first layer for improved accuracy. Each channel operates independently with its weights and accumulators, performing asynchronous MAC and synchronous thresholding in parallel. After the thresholding, one channel is activated at a time, and the output activations of the different channels are read sequentially utilizing the asynchronous Address-Event Representation (AER) read scheme \cite{aer_ref1, DVS_ref1} similar to the neuromorphic vision sensors. The output activation map is determined by kernel size and stride; hence, it is smaller than the raw sensor array. In addition, it eliminates the need for an extra bit to indicate event polarity (ON or OFF); hence, it reduces the number of required communicated off-chip address bits. Upon spike generation, a reset phase is executed for both positive and negative accumulators in the channel, involving a reset current flow in the negative direction to move the DW to the leftmost position.

\subsection{In-situ Multi-bit Weights}

\begin{figure}[!t]
\centering
\includegraphics[width=1\linewidth]{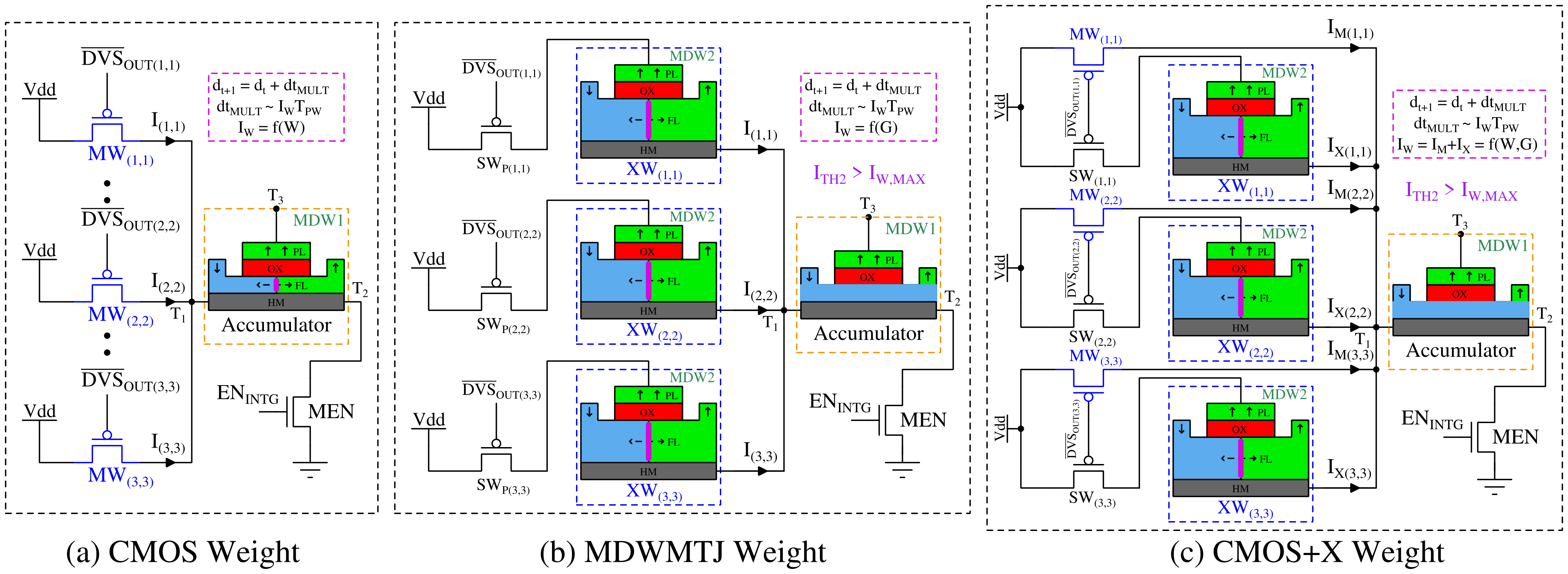}
\caption{Embedded in-situ multi-bit (a) CMOS-based, (b) MDWMTJ-based, and (c) CMOS+X-based weight implementation. Transistors \si{MW_{x,y}}, MDWMTJs \si{XW_{x,y}} and transistors \si{MW_{x,y}} + MDWMTJs \si{XW_{x,y}} represent the weights in (a), (b), and (c), respectively, where, (x,y) = (1,1), (1,2), ... (3,3), considering a kernel size of 3\si{\times}3.}
\label{weight_arch}
\end{figure}

Figure \ref{weight_arch} presents three configurations: CMOS, MDWMTJ, and Hybrid CMOS+X weights to perform the MAC operation. In the CMOS-based implementation, weights (e.g., \si{MW_{(1,1)}}, \si{MW_{(2,2)}}, \si{MW_{(3,3)}}) are represented by the transistor's width. The MDWMTJ-based configuration employs MDWMTJ as the weight (e.g., \si{XW_{(1,1)}}, \si{XW_{(2,2)}}, \si{XW_{(3,3)}}), and the resistance state (DW position) of the MDWMTJ dictates the weight value. The threshold current of the weight MDWMTJs needs to be higher (\si{I_{TH2}} > \si{I_{W,MAX}}) to prevent the weight current from triggering the writing process into the weight MDWMTJ during MAC operations. The thickness of the FL can modulate the threshold current of the MDWMTJ \cite{sot_Ith}. Though MDWMTJ-based configuration offers programmability, it exhibits a smaller dynamic range (due to low TMR \cite{mdw_exp2}, TMR = 200\% used in this work) than CMOS-based weights, which are fixed during the fabrication steps. To overcome these limitations and achieve versatility, we propose a hybrid CMOS+X configuration, combining transistor width and MDWMTJ resistance for adjustable write currents. This hybrid approach provides a broad dynamic range akin to CMOS-based weights and modest tunability (30-40\% for most weight values) like the MDW-based weights, requiring device-algorithm co-design optimization for effective implementation across various applications. Positive (negative) weights are connected to the positive (negative) accumulator MDWMTJs. A larger weight value corresponds to a wider transistor or/and a lower MDWMTJ resistance, resulting in a higher current flowing through the HM of the accumulator MDWMTJs during an event spike. The DW motion, triggered by the input spike, moves from left to right and accumulates the MAC results throughout the integration time. Due to non-volatility, the accumulator MDWMTJs retain the previous membrane potential (DW position) for subsequent integration times, a behavior important for achieving a high classification accuracy.

\subsection{Analog Convolution Operation} 

\begin{figure}[!t]
\centering
\includegraphics[width=0.75\linewidth]{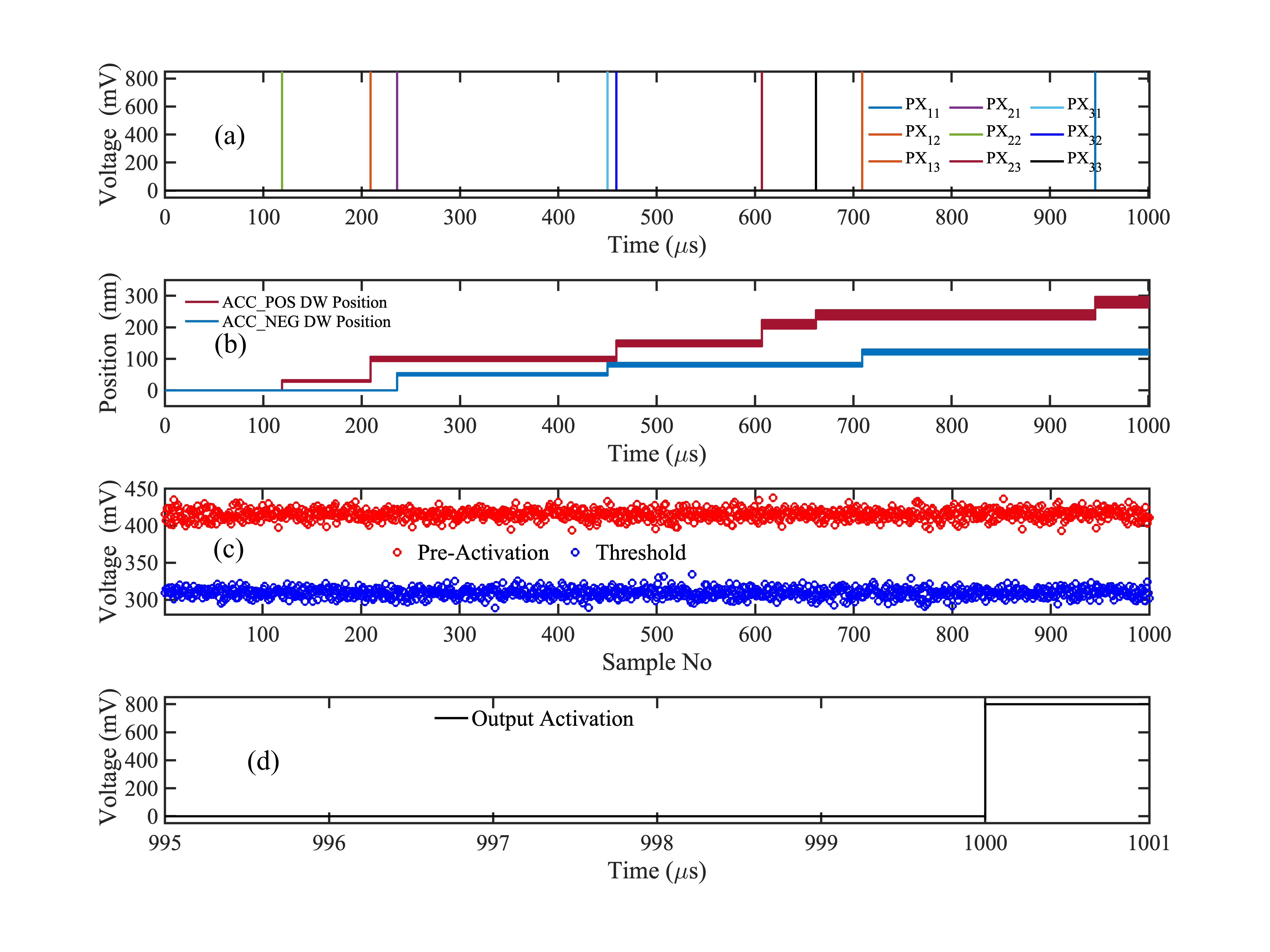}
\caption{1000 monte-carlo simulations of a random convolution operation with output activation spike simulated on GF 22nm FD-SOI node.}
\label{kernel_mc}
\end{figure}

Figure \ref{kernel_mc} presents an HSpice simulation of our hybrid CMOS+X approach based on asynchronous MAC computations and synchronous output activation using GF22nm FD-SOI technology. The simulation considers a 3\si{\times}3 kernel size, random positive and negative weights, and random event timings and includes transistor 3-sigma variation, 10\% Gaussian jitter in the write pulse, and 20\% resistance variations in MDWMTJs. Subplot (b) illustrates the rightward movement of positive and negative accumulators during event spikes, as seen in subplot (a). Despite using a 1 ms integration time for efficient 1000 monte-carlo simulations, the MDWMTJ's non-volatility allows longer state retention, with cumulative CMOS leakage staying below the accumulator MDWMTJ's threshold current, preventing any DW movement. After 1 ms of integration time, the reasonable sense margin between the final pre-activation and threshold voltage for 1000 samples (subplot (c)) leads to an output activation spike for the next layer (subplot (d)).

\begin{figure}[!b]
\centering
\includegraphics[width=0.75\linewidth]{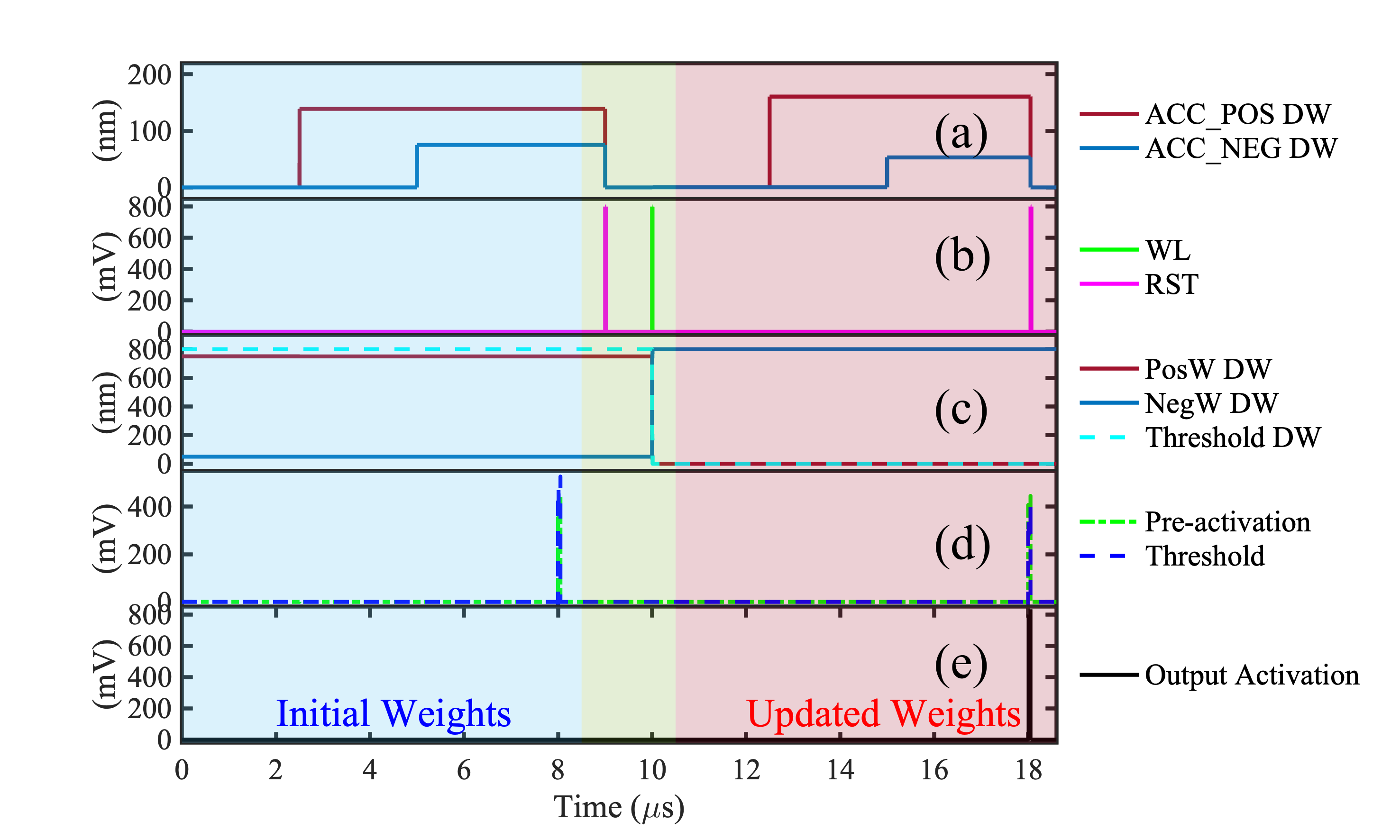}
\caption{Weight and threshold voltage reprogrammability simulation for two different applications.}
\label{reconfig}
\end{figure}

\subsection{Reprogrammability of Weight and Neuron's Threshold} 

In Figure \ref{reconfig}, we showcase the reprogrammability of weights and threshold voltage in our hybrid CMOS+X configuration, where we have utilized both CMOS and MDWMTJ as weights. The blue and red shaded regions represent MAC and thresholding operations for initial and updated setups, while the green region indicates the programmability of weights and threshold voltage, along with accumulator reset. One positive and one negative weight were randomly selected for this test. The transistor width is fixed in the CMOS+X hybrid configuration, and MDWMTJ resistance can be tuned for weight modulation. Initially (blue-shaded region), the MDWMTJ of the positive (negative) weight is set to a higher (lower) resistance state by programming the DW position. Depending on input event spikes, the DW of the positive (ACC\_POS DW in subplot (a)) and negative (ACC\_NEG DW in subplot (a)) accumulators moves rightward from their reset state. From subplots (d) and (e), it can be observed that the pre-activation voltage is smaller than the threshold voltage for the initial weight setup, resulting in an output activation of 0. The reset and programmable features are then demonstrated. The accumulators are reset, and the DW positions of the positive and negative MDWMTJ weights are programmed to different states within the tunable range (subplot (c)).
Additionally, the DW position of the reference voltage generation is programmed to a lower resistance state to achieve a smaller neuron threshold. Due to the updated weights and threshold voltage, the MAC and threshold results differ from the previous application. The difference between the positive and negative accumulator has increased, leading to a higher pre-activation voltage, surpassing the lowered threshold voltage, and resulting in an output activation spike. 

\section{Device-Circuit-Algorithm Co-design}

\begin{figure}[!b]
\centering
\subfloat[CMOS-based Weight]{\includegraphics[width=0.40\linewidth]{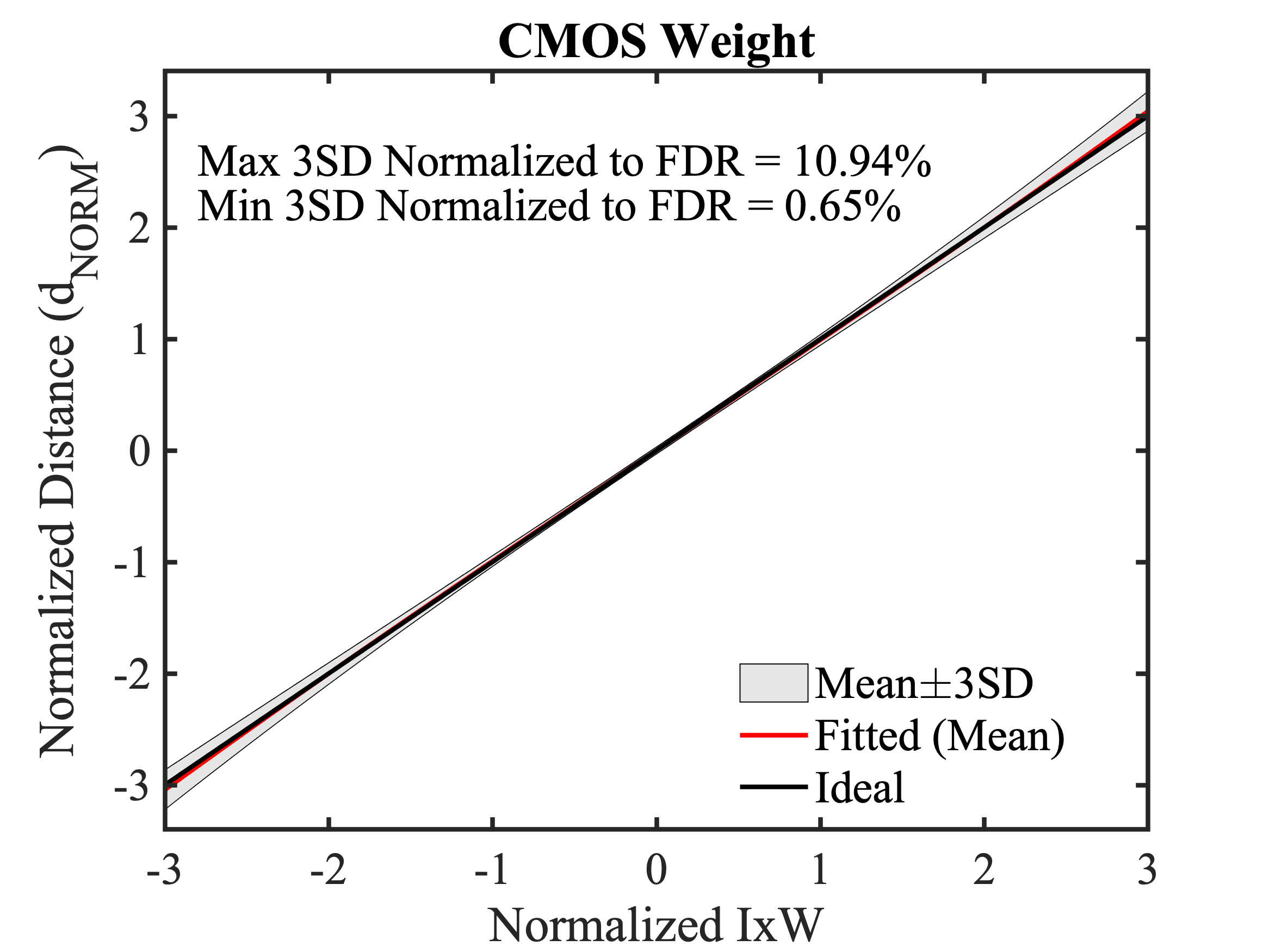}}
\subfloat[MDWMTJ-based Weight]{\includegraphics[width=0.40\linewidth]{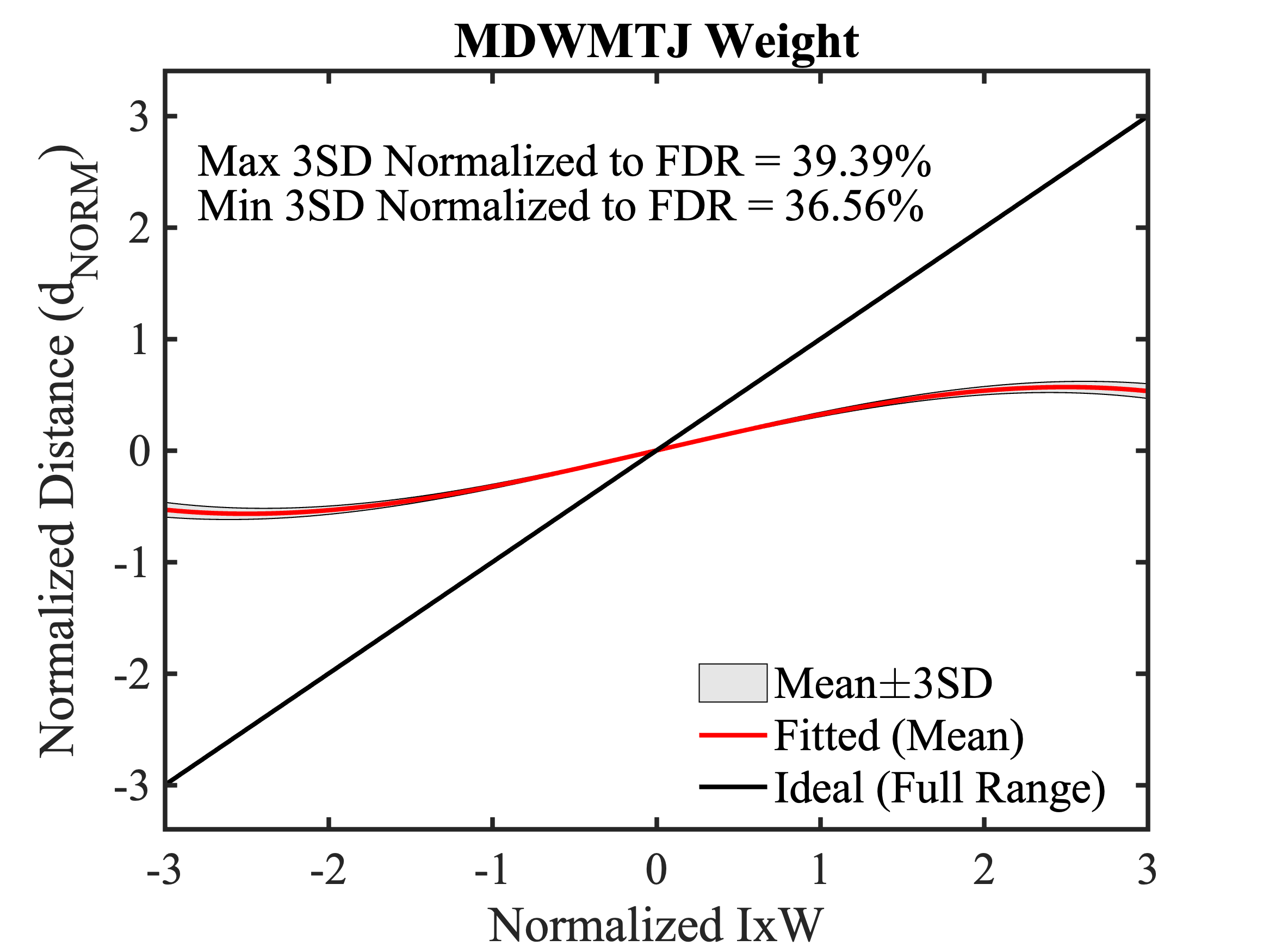}} \\
\subfloat[CMOS+X Weight]{\includegraphics[width=0.40\linewidth]{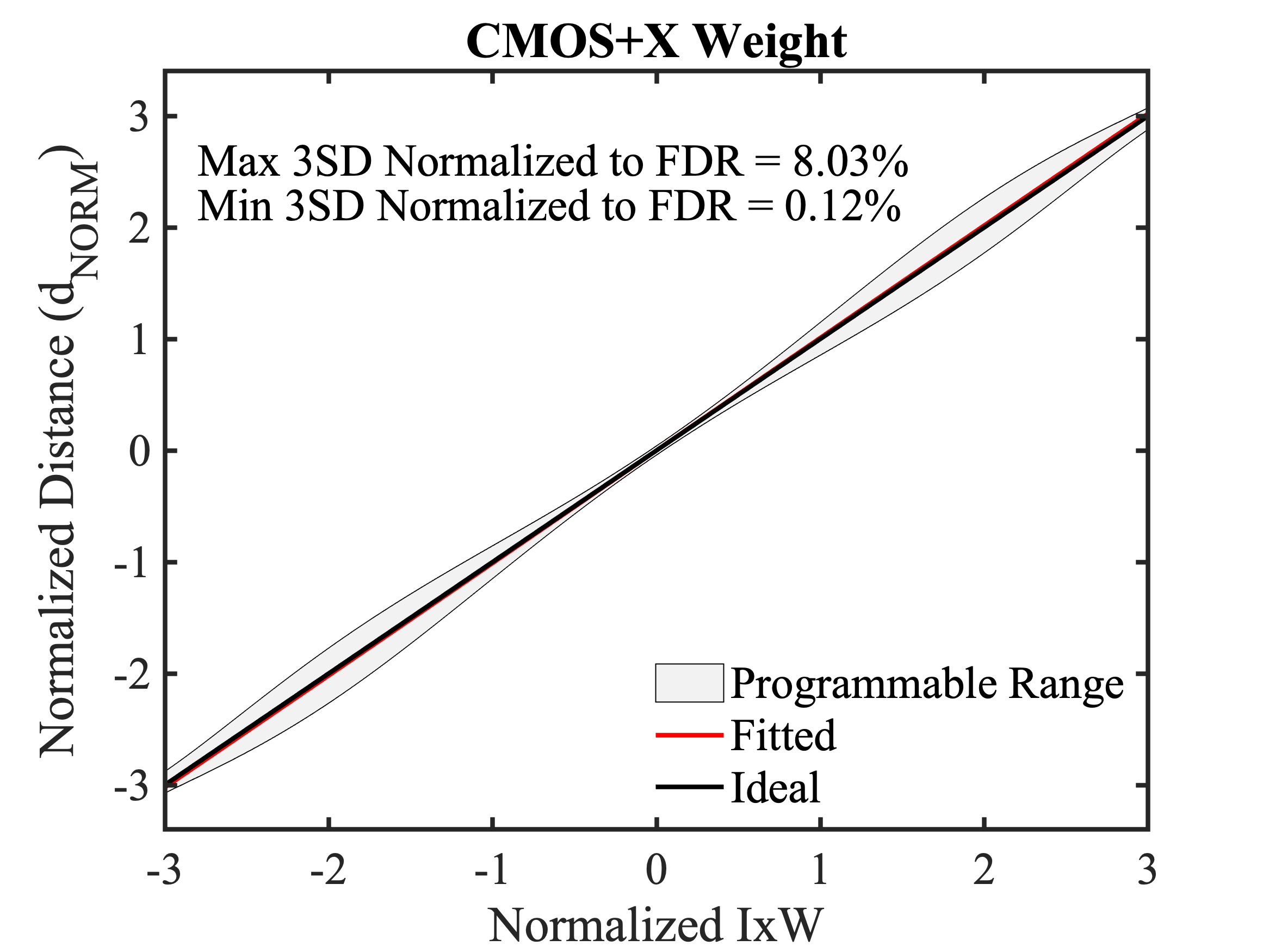}}
\subfloat[Read Voltage]{\includegraphics[width=0.40\linewidth]{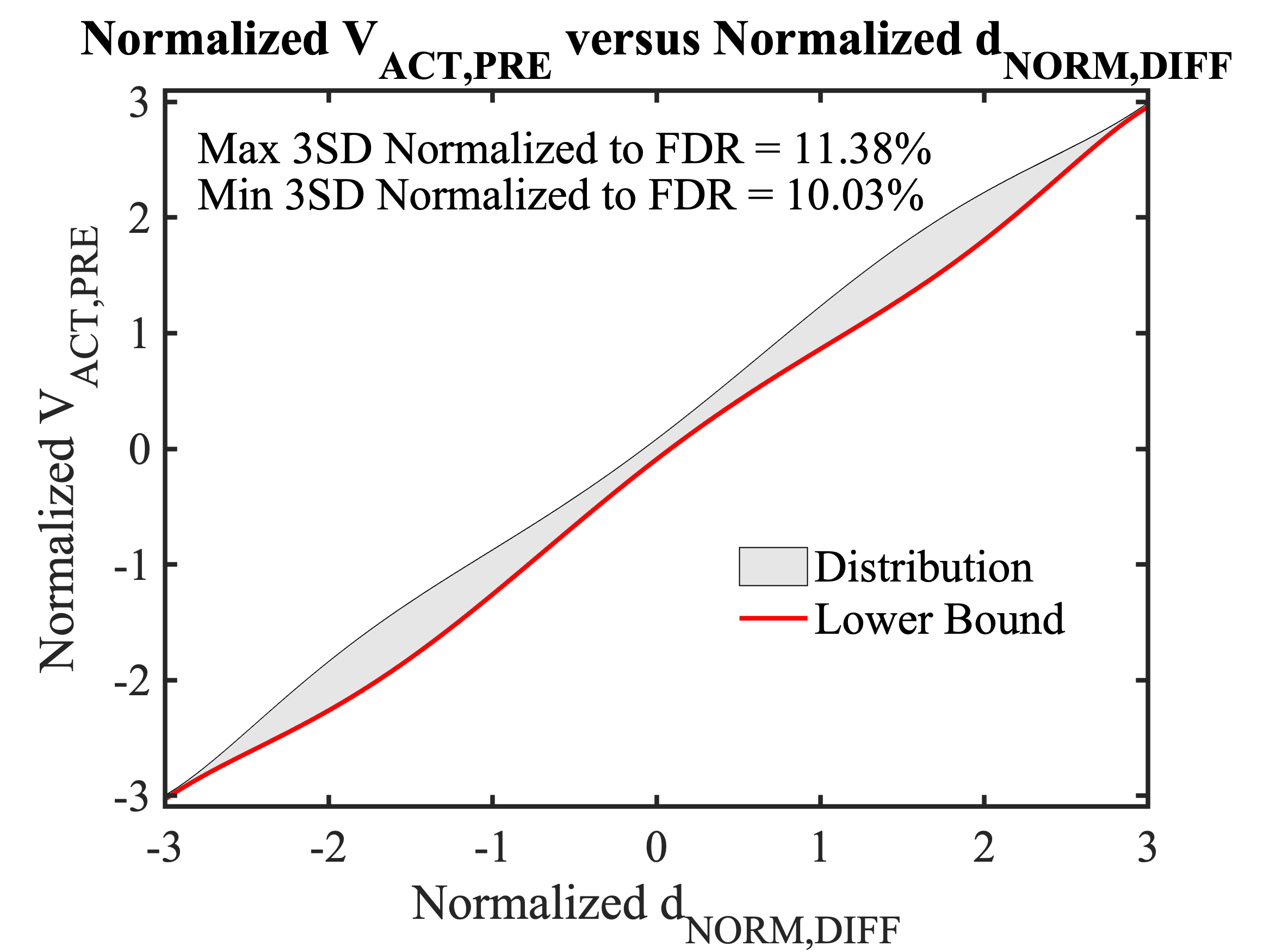}}
\caption{Normalized DW position (\si{d_{NORM}}) versus the normalized input activation\si{\times}weight characteristics for (a) CMOS, (b) MDWMTJ, and (c) CMOS+X weight configurations considering non-linearity and process variations. (d) Normalized pre-activation voltage (\si{V_{ACT,PRE}}) versus the difference in MAC results (\si{d_{NORM,DIFF}}) of the accumulators.}
\label{scatter_plot}
\end{figure}

In this section, we detail the implementation of our algorithmic framework on the proposed neuromorphic CMOS+X P\si{^2}M architecture, considering both device and circuit constraints. We address the non-ideal, non-linear attributes of MDWMTJ and transistors, incorporating process variations to account for potential resistance and current deviations in our CMOS+X system. Through extensive HSpice simulations on the GF22nm FD-SOI node, we map and integrate our circuit characteristics into our spiking CNN framework, utilizing custom functions that adapt to circuit and device non-linearity, non-ideality, and variations. 

\textbf{MAC output Modeling:} We encode spiking CNN filter weights as the current through the HM of accumulator MDWMTJs, adjusting the weight transistor's width or/and weight MDWMTJ's resistance (DW position). Domain wall motion depends on the write current pulse, exhibiting a non-linear relationship with the current (Figure \ref{device_modeling}(b)). Moreover, the non-linear and voltage-dependent write current by the transistor or/and MDWMTJ is also affected by process variations, fabrication uncertainty, and noise. Simulating for various input spikes and weight combinations, accounting for 3-sigma variation for GF22nm FD-SOI devices, 20\% MDWMTJ resistance variation, and 10\% clock jitter in the write pulse, our HSpice results for CMOS-based, MDWMTJ-based, and CMOS+X-based configurations (Figures \ref{scatter_plot}(a), (b), and (c), respectively) are modeled using a behavioral curve-fitting function (\si{f_1}). Due to the low dynamic range (DR), the MDWMTJ-based weight exhibits higher worst-case variation (39.39\%) compared to the CMOS (10.94\%) and CMOS+X (8.03\%) weights normalized to full DR. The CMOS+X configuration reports a 30-40\% normalized programmable range for most weights.

\textbf{Pre-activation result modeling:} After accumulating the DW motion for a fixed integration time, the pre-activation voltage is generated from the series division of the negative and positive accumulator MDWMTJs, where the subtracted outcome (positive MAC output - negative MAC output) non-linearly influences the output voltage of the series divider. Due to the non-linearity and voltage-dependent resistance of the MDWMTJ, identical differences in MAC results may generate different pre-activation voltages. Addressing this non-linearity through extensive HSpice simulations for various combinations (depicted in Figure \ref{scatter_plot}(d)), we model the worst-case scenario (lower bound) that exhibits a maximum of 11.38\% variations, utilizing a behavioral curve-fitting function (\si{f_2}) for the normalized pre-activation voltage versus the normalized distance difference of the accumulator MDWMTJs.

\textbf{Circuit-Algorithm co-design optimization:} Our algorithmic framework generates random Gaussian sample values based on mean and standard deviation results from HSpice simulations to address process variation effects. The accumulated DW position for each pixel's event spike is calculated using the \si{f_1} function incorporating hardware's non-ideality, aggregated throughout the integration time, to determine MAC results for positive and negative weights. The normalized pre-activation voltage is computed through the \si{f_2} function to incorporate the non-linearity and variations of the read circuits. The worst-case scenario (lower bound) is used for pre-activation voltage calculation to ensure accurate output spike generation. This algorithmic framework, employing custom functions \si{f_1} and \si{f_2}, optimizes spiking CNN training for event-driven neuromorphic datasets. In addition, our algorithmic framework is also optimized utilizing 32 channels with stride 2 for the first layer, ensuring no area overhead for our P\si{^2}M core.

\textbf{Device-Algorithm co-design optimization:} Due to the limited TMR, our hybrid CMOS+X P\si{^2}M configuration offers a programmability range 30-40\% from its median value. We employ weight tunability restrictions in our algorithmic framework to accommodate diverse applications, enabling our hardware to cater to various applications. Our approach involves initially training weights for one application and subsequently retraining them, considering the normalized weight tunability range derived from HSpice simulation results for another application. Additionally, we leverage kernel-dependent adjustments to neuron thresholds to optimize the algorithmic accuracy.

\section{Experimental Results}

In this work, we explore and benchmark our proposed neuromorphic CMOS+X P\si{^2}M solution utilizing the event-driven neuromorphic tasks, specifically classifying video samples captured by NVS cameras. The evaluation is conducted on three widely used neuromorphic benchmark datasets: NMNIST \cite{nmnist}, CIFAR10-DVS \cite{cifar10dvs}, and IBM DVS128-Gesture \cite{dvs-gesture}. The Spikingjelly package \cite{SpikingJelly} is employed to process and integrate data into fixed time intervals, and a 9:1 train-validation split is applied for these datasets. The spiking CNN architecture consists of four convolutional layers, succeeded by two linear layers featuring 512 and 10 neurons, respectively. Following each convolutional layer, there is a sequence of a batch normalization layer, a spiking LIF layer, and a max pooling layer.

\begin{table}[!t]
\caption{Evaluation of our P\si{^2}M approach modeled using a custom first layer for NVS datasets, where the noise denotes the variation in the custom functions \si{f_1} and \si{f_2}.}
\label{tab:acc}
\begin{center}
\resizebox{0.50\columnwidth}{!}{%
\begin{tabular}{cccc}
Dataset & Custom Func. & Noise (\%) & Accuracy (\%) \\ 
\hline
\multirow{5}{*}{NMNIST} &  No (Baseline) & 0 & 98.10 \\
\cline{2-4}
& Yes & 0 & 98.04  \\
\cline{2-4}
& Yes & 10 & 97.82  \\
\cline{2-4}
& Yes & 20 & 97.71  \\
\cline{2-4}
& Yes & 40 & 95.42 \\
\hline
\multirow{5}{*}{CIFAR10-DVS} &  No (Baseline) & 0 & 80.72 \\
\cline{2-4}
& Yes & 0 & 80.25  \\
\cline{2-4}
& Yes & 10 & 79.17  \\
\cline{2-4}
& Yes & 20 & 77.80  \\
\cline{2-4}
& Yes & 40 & 65.58 \\
\hline
\multirow{5}{*}{DVS128-Gesture} &  No (Baseline) & 0 & 97.22 \\
\cline{2-4}
& Yes & 0 & 96.53  \\
\cline{2-4}
& Yes & 10 & 95.99  \\
\cline{2-4}
& Yes & 20 & 95.15  \\
\cline{2-4}
& Yes & 40 & 87.25 \\
\hline
\end{tabular}}%
\vspace{-0.2in}
\end{center}
\end{table}

\begin{table}[!t]
\caption{Efficacy of the reprogrammability of our proposed CMOS+X approach.}
\label{tab:reconfig}
\begin{center}
\resizebox{0.50\columnwidth}{!}{%
\begin{tabular}{cccc}
Train & Evaluate & Fine-tune & Accuracy (\%) \\ 
\hline
\multirow{3}{*}{CIFAR10-DVS} & \multirow{3}{*}{DVS128-Gesture} & No & 9.12 \\
\cline{3-4}
 &   &  Except $1^{st}$ layer & 94.96 \\
\cline{3-4}
&  & Yes & 96.51 \\
\hline
\multirow{3}{*}{DVS128-Gesture} & \multirow{3}{*}{CIFAR10-DVS} & No & 10.10 \\
\cline{3-4}
 &   & Except $1^{st}$ layer & 76.73 \\
\cline{3-4}
&  & Yes & 80.14 \\
\hline
\end{tabular}}%
\vspace{-0.2in}
\end{center}
\end{table}

\subsection{Classification Accuracy \& Programmablity}

Our spiking CNNs with in-pixel processing achieve test accuracy comparable to SNNs processed with traditional digital hardware (baseline), as demonstrated in Table \ref{tab:acc}. This success is attributed to the non-volatility of MDWMTJ and our algorithmic optimization considering variations. Additionally, our CMOS+X approach allows re-programmability, enabling post-fabrication tuning of the first layer weights in our P\si{^2}M hardware for diverse neuromorphic applications. As indicated in Table \ref{tab:reconfig}, fine-tuning, especially of the first layer, is essential for maintaining accuracy when transitioning from training on the DVS-CIFAR10 or IBM DVS128-Gesture datasets to other applications. The absence of this fine-tuning results in a significant accuracy drop.

\begin{figure}[!t]
\centering
\includegraphics[width=0.75\linewidth]{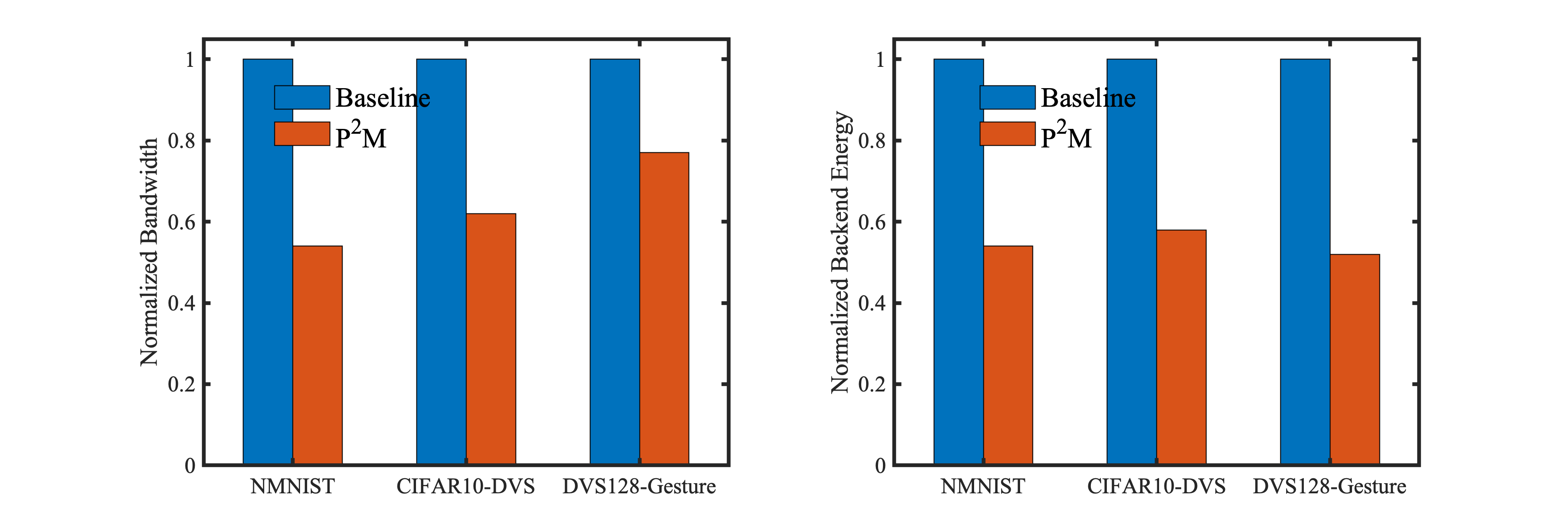}
\caption{Normalized bandwidth (left-subplot) and backend energy (right-subplot) for the baseline and P\si{^2}M-enabled spiking CNN for NVS datasets.}
\label{fig:p2m_savings}
\end{figure}

\subsection{Details of Bandwidth \& Energy Savings}

Fig. \ref{fig:p2m_savings} demonstrates the bandwidth and energy savings with our in-pixel processing approach for the three neuromorphic datasets. We calculate bandwidth as the ratio of average output activation spikes to input event spikes per input sample. To reduce the bandwidth, we employ the bit-level $\ell_1$ regularizer in the first layer, similar to \cite{sekikawa2023bitpruning}. The
normalized bandwidth reductions of our approach compared to the baseline, where the input spikes are directly sent to the backend processor, are 0.54, 0.62, and 0.77 for NMNIST, DVS-CIFAR10, and IBM DVS128-Gesture, respectively. The backend compute energy consumption is normalized with respect to the conventional backend processing (i.e., digital implementation). Due to the in-pixel processing of the first spiking CNN layer in the analog domain, our approach yields an average (across 3 NVS datasets) of 45.3\% lower backend energy. Note that the backend energy is determined by the number of spikes generated by each layer and the memory access of the weights and membrane potential, similar to \cite{sata}.

\section{Conclusion}

We have proposed and implemented a novel CMOS+X processing-in-pixel-in-memory paradigm for neuromorphic event-based sensors. To our knowledge, this is the first CMOS+X implementation proposal for in-pixel processing focusing on neuromorphic vision sensor applications. Leveraging the non-volatility and programmable features of the MDWMTJ and the high dynamic range achieved by the CMOS transistors,  our developed spiking CNN hardware can cater to various applications, yielding SOTA accuracy. In addition, we optimize and evaluate our system, incorporating device and circuit constraints into our algorithmic framework based on extensive HSpice simulations. Our neuromorphic CMOS+X P\si{^2}M-enabled spiking CNN model yields an accuracy of 97.82\%, 79.17\%, and 95.99\% on the NMNIST, CIFAR10-DVS, IBM DVS128-Gesture datasets, respectively and achieved an average of 45.3\% backend energy reduction compared to the conventional system.

\bibliographystyle{unsrtnat}
\bibliography{references}

\begin{thebibliography}{38}
\providecommand{\natexlab}[1]{#1}
\providecommand{\url}[1]{\texttt{#1}}
\expandafter\ifx\csname urlstyle\endcsname\relax
  \providecommand{\doi}[1]{doi: #1}\else
  \providecommand{\doi}{doi: \begingroup \urlstyle{rm}\Url}\fi

\bibitem[Chai(2020)]{system_bottleneck}
Yang Chai.
\newblock In-sensor computing for machine vision, 2020.

\bibitem[Eki et~al.(2021)]{near_sensor_3D_sony}
Ryoji Eki et~al.
\newblock A 1/2.3 inch 12.3 mpixel with on-chip 4.97 tops/w cnn processor back-illuminated stacked cmos image sensor.
\newblock In \emph{ISSCC 2021}, volume~64, pages 154--156. IEEE, 2021.

\bibitem[Lefebvre et~al.(2021)]{sleepspotter}
Martin Lefebvre et~al.
\newblock A 0.2-to-3.6 tops/w programmable convolutional imager soc with in-sensor current-domain ternary-weighted mac operations for feature extraction and region-of-interest detection.
\newblock In \emph{ISSCC 2021}, volume~64, pages 118--120. IEEE, 2021.

\bibitem[Tabrizchi et~al.(2023)]{inpix_conv}
Sepehr Tabrizchi et~al.
\newblock Appcip: Energy-efficient approximate convolution-in-pixel scheme for neural network acceleration.
\newblock \emph{IEEE Journal on Emerging and Selected Topics in Circuits and Systems}, 13\penalty0 (1):\penalty0 225--236, 2023.

\bibitem[Datta et~al.(2022{\natexlab{a}})]{aps_p2m}
Gourav Datta et~al.
\newblock A processing-in-pixel-in-memory paradigm for resource-constrained tinyml applications.
\newblock \emph{Scientific Reports}, 12, 2022{\natexlab{a}}.

\bibitem[Hsu et~al.(2022)]{mixed_mode_ivs}
Tzu-Hsiang Hsu et~al.
\newblock A 0.8 v intelligent vision sensor with tiny convolutional neural network and programmable weights using mixed-mode processing-in-sensor technique for image classification.
\newblock In \emph{ISSCC 2022}, volume~65, pages 1--3. IEEE, 2022.

\bibitem[Lichtsteiner et~al.(2008)]{DVS_ref1}
Patrick Lichtsteiner et~al.
\newblock A 128x128 120 db 15 $\mu$s latency asynchronous temporal contrast vision sensor.
\newblock \emph{IEEE JSSC}, 43\penalty0 (2):\penalty0 566--576, 2008.

\bibitem[Le{\~n}ero-Bardallo et~al.(2011)]{DVS_ref2}
Juan~Antonio Le{\~n}ero-Bardallo et~al.
\newblock A 3.6 $\mu$s latency asynchronous frame-free event-driven dynamic-vision-sensor.
\newblock \emph{IEEE JSSC}, 46\penalty0 (6):\penalty0 1443--1455, 2011.

\bibitem[Chen et~al.(2020)]{DVS_auto_driving}
Guang Chen et~al.
\newblock Event-based neuromorphic vision for autonomous driving: A paradigm shift for bio-inspired visual sensing and perception.
\newblock \emph{IEEE Signal Processing Magazine}, 37\penalty0 (4):\penalty0 34--49, 2020.

\bibitem[Nguyen et~al.(2019)]{DVS_pose}
Anh Nguyen et~al.
\newblock Real-time 6dof pose relocalization for event cameras with stacked spatial lstm networks.
\newblock In \emph{CVPR}, pages 0--0, 2019.

\bibitem[Maqueda et~al.(2018)]{DVS_steering}
Ana~I Maqueda et~al.
\newblock Event-based vision meets deep learning on steering prediction for self-driving cars.
\newblock In \emph{IEEE CVPR}, pages 5419--5427, 2018.

\bibitem[Datta et~al.(2022{\natexlab{b}})]{spiking_cnn}
Gourav Datta et~al.
\newblock Can deep neural networks be converted to ultra low-latency spiking neural networks?
\newblock In \emph{DATE 2022}, volume~1, pages 718--723, 2022{\natexlab{b}}.

\bibitem[Song et~al.(2022)]{reconfig_inpix}
Ruibing Song et~al.
\newblock A reconfigurable convolution-in-pixel cmos image sensor architecture.
\newblock \emph{IEEE Transactions on Circuits and Systems for Video Technology}, 32\penalty0 (10):\penalty0 7212--7225, 2022.

\bibitem[Zhang et~al.(2022)]{inmemory_nvs}
Xueyong Zhang et~al.
\newblock A 915--1220 tops/w, 976--1301 gops hybrid in-memory computing based always-on image processing for neuromorphic vision sensors.
\newblock \emph{IEEE JSSC}, 58\penalty0 (3):\penalty0 589--599, 2022.

\bibitem[Hsu et~al.(2020)]{current_cap_integ}
Tzu-Hsiang Hsu et~al.
\newblock A 0.5-v real-time computational cmos image sensor with programmable kernel for feature extraction.
\newblock \emph{IEEE JSSC}, 56\penalty0 (5):\penalty0 1588--1596, 2020.

\bibitem[Kaiser et~al.(2023)Kaiser, Datta, Wang, Jacob, Beerel, and Jaiswal]{dvsp2m}
Md~Abdullah-Al Kaiser, Gourav Datta, Zixu Wang, Ajey~P Jacob, Peter~A Beerel, and Akhilesh~R Jaiswal.
\newblock Neuromorphic-p2m: processing-in-pixel-in-memory paradigm for neuromorphic image sensors.
\newblock \emph{Frontiers in Neuroinformatics}, 17:\penalty0 1144301, 2023.

\bibitem[Roy et~al.(2020)]{inmem_emerging}
Kaushik Roy et~al.
\newblock In-memory computing in emerging memory technologies for machine learning: An overview.
\newblock In \emph{DAC 2020}, pages 1--6. IEEE, 2020.

\bibitem[Ankit et~al.(2017)]{inmem_rram}
Aayush Ankit et~al.
\newblock Resparc: A reconfigurable and energy-efficient architecture with memristive crossbars for deep spiking neural networks.
\newblock In \emph{DAC 2017}, pages 1--6, 2017.

\bibitem[Sengupta et~al.(2016)]{mdw_ann}
Abhronil Sengupta et~al.
\newblock Proposal for an all-spin artificial neural network: Emulating neural and synaptic functionalities through domain wall motion in ferromagnets.
\newblock \emph{TBioCAS}, 10\penalty0 (6):\penalty0 1152--1160, 2016.

\bibitem[Leonard et~al.(2022)]{mdw_exp1}
Thomas Leonard et~al.
\newblock Shape-dependent multi-weight magnetic artificial synapses for neuromorphic computing.
\newblock \emph{Advanced Electronic Materials}, 8\penalty0 (12):\penalty0 2200563, 2022.

\bibitem[Alamdar et~al.(2021)]{mdw_exp2}
Mahshid Alamdar et~al.
\newblock Domain wall-magnetic tunnel junction spin-orbit torque devices and circuits for in-memory computing.
\newblock \emph{APL}, 118\penalty0 (11), 2021.

\bibitem[Ikeda et~al.(2008)]{tmr_600p}
S~Ikeda et~al.
\newblock Tunnel magnetoresistance of 604\% at 300k by suppression of ta diffusion in cofeb/ mgo/ cofeb pseudo-spin-valves annealed at high temperature.
\newblock \emph{APL}, 93\penalty0 (8), 2008.

\bibitem[Hu et~al.(2019)]{mdw_logic1}
Xuan Hu et~al.
\newblock Spice-only model for spin-transfer torque domain wall mtj logic.
\newblock \emph{IEEE TED}, 66\penalty0 (6):\penalty0 2817--2821, 2019.

\bibitem[Wang et~al.(2020)]{mdwmtj_ref1}
Chao Wang et~al.
\newblock Compact model of dzyaloshinskii domain wall motion-based mtj for spin neural networks.
\newblock \emph{IEEE TED}, 67\penalty0 (6):\penalty0 2621--2626, 2020.

\bibitem[Wang et~al.(2021)]{mdwmtj_ref2}
Manman Wang et~al.
\newblock Compact model of domain wall mtj driven by spin-orbit torque and dzyaloshinskii--moriya interaction.
\newblock \emph{IEEE Transactions on Magnetics}, 58\penalty0 (8):\penalty0 1--5, 2021.

\bibitem[Martinez et~al.(2014)]{dmi_ref1}
Eduardo Martinez et~al.
\newblock Current-driven dynamics of dzyaloshinskii domain walls in the presence of in-plane fields: Full micromagnetic and one-dimensional analysis.
\newblock \emph{Journal of Applied Physics}, 115\penalty0 (21), 2014.

\bibitem[Luo et~al.(2021)]{mdw_integrator}
Shijiang Luo et~al.
\newblock Integrator based on current-controlled magnetic domain wall.
\newblock \emph{APL}, 118\penalty0 (5), 2021.

\bibitem[Fong et~al.(2011)]{mdw_negf}
Xuanyao Fong et~al.
\newblock Knack: A hybrid spin-charge mixed-mode simulator for evaluating different genres of spin-transfer torque mram bit-cells.
\newblock In \emph{2011 International Conference on Simulation of Semiconductor Processes and Devices}, pages 51--54. IEEE, 2011.

\bibitem[Miura et~al.(2019)]{3D_ref1}
Tsukasa Miura et~al.
\newblock A 6.9 $\mu$m pixel-pitch 3d stacked global shutter cmos image sensor with 3m cu-cu connections.
\newblock In \emph{3DIC 2019}, pages 1--2. IEEE, 2019.

\bibitem[Kagawa et~al.(2020)]{3D_ref2}
Y~Kagawa et~al.
\newblock Impacts of misalignment on 1$\mu$m pitch cu-cu hybrid bonding.
\newblock In \emph{IITC 2020}, pages 148--150. IEEE, 2020.

\bibitem[Boahen(2004)]{aer_ref1}
Kwabena~A Boahen.
\newblock A burst-mode word-serial address-event link-i: Transmitter design.
\newblock \emph{IEEE TCAS-I}, 51\penalty0 (7):\penalty0 1269--1280, 2004.

\bibitem[Zhu et~al.(2020)]{sot_Ith}
Daoqian Zhu et~al.
\newblock Threshold current density for perpendicular magnetization switching through spin-orbit torque.
\newblock \emph{Physical Review Applied}, 13\penalty0 (4):\penalty0 044078, 2020.

\bibitem[Orchard et~al.(2015)]{nmnist}
Garrick Orchard et~al.
\newblock Converting static image datasets to spiking neuromorphic datasets using saccades.
\newblock \emph{Frontiers in Neuroscience}, 9, 2015.
\newblock URL \url{https://www.frontiersin.org/articles/10.3389/fnins.2015.00437}.

\bibitem[Li et~al.(2017)]{cifar10dvs}
Hongmin Li et~al.
\newblock Cifar10-dvs: An event-stream dataset for object classification.
\newblock \emph{Frontiers in Neuroscience}, 11, 2017.
\newblock URL \url{https://www.frontiersin.org/articles/10.3389/fnins.2017.00309}.

\bibitem[Amir et~al.(2017)]{dvs-gesture}
Arnon Amir et~al.
\newblock A low power, fully event-based gesture recognition system.
\newblock In \emph{CVPR 2017}, volume~1, pages 7388--7397, 2017.

\bibitem[Fang et~al.(2020)]{SpikingJelly}
Wei Fang et~al.
\newblock Spikingjelly, 2020.

\bibitem[Sekikawa et~al.(2023)]{sekikawa2023bitpruning}
Yusuke Sekikawa et~al.
\newblock Bit-pruning: A sparse multiplication-less dot-product.
\newblock In \emph{The Eleventh International Conference on Learning Representations}, 2023.
\newblock URL \url{https://openreview.net/forum?id=YUDiZcZTI8}.

\bibitem[Yin et~al.(2023)]{sata}
R~Yin et~al.
\newblock Sata: Sparsity-aware training accelerator for spiking neural networks.
\newblock \emph{IEEE Transactions on Computer-Aided Design of Integrated Circuits and Systems}, 42\penalty0 (6):\penalty0 1926--1938, 2023.

\end{thebibliography}

\end{document}